\documentclass[aip,apl,amsmath,amssymb,reprint]{revtex4-1}
\usepackage{bm,graphicx}
\usepackage{xcolor}

\begin{document}

\title{On quantum confinement without barriers}

\author{I.A. Kokurin}
\email[E-mail:]{ivan.a.kokurin@gmail.com} \affiliation{Ioffe
Institute, 194021 St. Petersburg, Russia}

\begin{abstract}
Usually in semiconductor heterostructures, such as quantum wells,
the quantum confinement is realized by means of heterobarriers,
which are simply the band offsets in adjacent materials. Here the
arguments are provided, that the energy dispersions in materials (in
simple case determined by effective masses) are important as well.
Using a simple model system, such as quantum well with an additional
heterojunction, it is shown the predominant localization of the
carrier density in the regions with higher effective mass. The
possibility of different spatial localization of wave function for
different subbands is predicted. This can be applied in
semiconductor structures such as quantum cascade lasers.
\end{abstract}

\maketitle

The modern semiconductor physics usually deals with semiconductor
heterostructures. Until now, the effective mass and envelope
function approximations remain the main methods for calculating the
subband spectrum of semiconductor heterostructures such as quantum
wells (QWs). The presence of sharp heterointerfaces is usually
assumed, which allows using a standard approach based on applying
boundary conditions (BCs) to the envelopes at heterointerfaces (see,
for instance, Ref.~\cite{Bastard1988}).

The band offsets at the heterointerfaces serve as barriers, the
presence of which leads to the quantum confinement. The charge
carriers are spatially localized in the regions, where the
conduction band bottom is located lower (the valence band top is
higher). The difference in energy-band parameters (e.g., the
effective masses in the simple conduction band) on opposite sides of
the heterointerface is usually perceived as a certain complexity for
calculations, which does not carry anything new for the effect of
quantum confinement.

The idea that the wave function of a particle is localized in the
region where its total energy is minimal seems quite obvious. At the
same time, the possibility of localizing a quantum particle in the
region of a potential barrier looks like an utopia. However, for
charge carriers in a semiconductor heterostructure, whose effective
mass (or other band parameters) can be considered spatially
variable, the localization of wave functions in regions with higher
``potential'' energy due to a decrease in ``kinetic'' energy seems
realistic (at least for some states).

Here, using a simple structured QW (SQW) model, this possibility
will be demonstrated. The opportunity of localizing the wave
function in different spatial regions of QW for different states
appears interesting for potential applications. For example, in SQW
structures it is possible to control the efficiency of optical
transitions and obtain very long radiative lifetimes.

\begin{figure}
\includegraphics{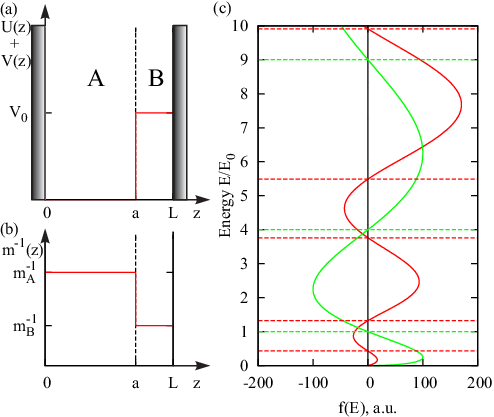}
\caption{\label{fig01} (a) The SQW potential profile $U(z)+V(z)$ is
schematically depicted. The shaded regions ($z<0$ and $z>L$) are
impenetrable to the charge carrier. The regions A ($0<z<a$) and B
($a<z<L$) are designated. (b) Spatial profile of the inverse
effective mass $m^{-1}(z)$ in SQW. (c) Function $f(E)$ (solid red
line) at $k=0$ and $m_B/m_A=10$. Intersection with vertical axis
[$f(E)=0$] yields the subband energies $E_n(0)$ (horizontal red
dashed lines). The function $f(E)$ at $m_A=m_B$ (solid green line)
is presented for comparison. Corresponding energy levels (horizontal
green dashed lines) are $E_n=E_0n^2$, $n=1,2,3,...$.}
\end{figure}

Let us consider the model of SQW, which consist of four
semiconducting layers (three heterointerfaces). The two outer layers
are of wide-gap materials, so the electron wave function hardly
penetrates this layers, and zero BCs for electron envelopes used at
heterointerfaces $z=0$ and $z=L$. The inner region consist of two
materials A and B: there is an additional heterointerface at $z=a$
($0<a<L$). The band offset at $z=a$ is equal to $V_0$. The potential
profile of such a structure is presented in Fig.~\ref{fig01}a. The
difference in the effective masses $m_A$ and $m_B$ of materials A
and B is taken into account (see Fig.~\ref{fig01}b). For the sake of
simplicity a simple isotropic conduction band is considered in both
material A and material B. It should be remembered that, as a rule,
a larger band gap corresponds to a larger effective mass. Usually,
the ratio $m_B/m_A$ is not to large, e.g., for
GaAs/Al$_x$Ga$_{1-x}$As heterostructure ($x=0.3$) the band offset
$V_0$ is about 300 meV, and the effective masses are $m_A=0.067$ and
$m_B=0.092$~\cite{Vurgaftman2001} in units of the bare electron mass
$m_0$.

Let us first consider SQW without an additional barrier, $V_0=0$,
whereas the inequality of effective masses is preserved, $m_A\neq
m_B$. The generalization to the case $V_0\neq 0$ is elementary. The
standard approach requires the application of BCs at each (in our
case three) heterointefrace. In addition to zero BCs on the left and
right heterointerfaces, the standard BCs conserving the envelope
function $\psi(z)$ and its flux at inner heterointerface $z=a$ are
used
\begin{equation}
\label{BCs} \psi(a-0)=\psi(a+0),\;\;\;
\left.\frac{1}{m_A}\frac{d\psi}{dz}\right|_{a-0}=\left.\frac{1}{m_B}\frac{d\psi}{dz}\right|_{a+0}.
\end{equation}
Obviously, the explicit form of the model and the corresponding BCs
are not essential for the effect discussed here.

Applying above BCs to the envelope functions in region A and B (both
are linear combinations of sine and cosine functions) the following
system of equations is derived:

\[
C\cos(k_BL)+D\sin(k_BL)=0,
\]
\[
B\sin(k_Aa)=C\cos(k_Ba)+D\sin(k_Ba),
\]
\[
\frac{k_A}{m_A}B\cos(k_Aa)=-\frac{k_B}{m_B}C\sin(k_Ba)+\frac{k_B}{m_B}D\cos(k_Ba).
\]

Here $k_{A(B)}=\sqrt{2m_{A(B)}E/\hbar^2-k^2}$ and $\hbar
k=\hbar\sqrt{k_x^2+k_y^2}$ is the momentum in QW plane. The
solvability condition for this system, relative to coefficients $B$,
$C$ and $D$, is given by

\begin{equation}
\left|\begin{array}{ccc}
0 & \cos(k_BL) & \sin(k_BL)\\
\sin(k_Aa) & -\cos(k_Ba) & -\sin(k_Ba)\\
\frac{k_A}{m_A}\cos(k_Aa) & \frac{k_B}{m_B}\sin(k_Ba) &
-\frac{k_B}{m_B}\cos(k_Ba)
\end{array}\right|=0.
\end{equation}

Finally, it leads to equation $f(E)=0$ with
\begin{eqnarray}
\label{f_E} \nonumber
f(E)&=&\frac{k_A}{m_A}\cos(k_Aa)\sin[k_B(L-a)]\\
&+&\frac{k_B}{m_B}\sin(k_Aa)\cos[k_B(L-a)],
\end{eqnarray}
whose numerical solution allows to determine the subband energy
spectrum. One can see, that at $m_A=m_B$ we have
$f(E)=\frac{k_A}{m_A}\sin(k_AL)$. The latter leads to the standard
spectrum $E_n=E_0n^2+\frac{\hbar^2k^2}{2m_A}$
($E_0=\frac{\pi^2\hbar^2}{2m_AL^2}$) of an electron in QW with
hard-wall boundaries.

The numerical solution of equation $f(E)=0$ with $f(E)$ of form
(\ref{f_E}) with fixed $k$ is easily solved by means of numerical
methods. This allows to find the subband spectrum $E_i(k)$
($i=1,2,3,...$). The dependence $f(E)$ at $k=0$ is plotted in
Fig.~\ref{fig01}c and the graphical solution (intersection with the
vertical axis) is presented for clarity. The corresponding wave
functions $\psi_i(z;k)$ can be found through the coefficients
$B(k)$, $C(k)$, $D(k)$, which, in turn, found from the BC system of
equations and the normalization condition.

Let us consider another way to solve the spectral problem. It is
based on the numerical diagonalization of the matrix Hamiltonian
written in the appropriate basis. The similar approach was applied
to calculate the subband spectrum of nanowire-based axial
heterostructures~\cite{Rudakov2019}. The general form of the
effective mass Hamiltonian is given by
\begin{equation}
\label{Hamiltonian}
H=-\frac{\hbar^2}{2}\nabla\frac{1}{m(z)}\nabla+U(z)+V(z).
\end{equation}
Here $U(z)$ is the hard-wall potential, $V(z)$ is the additional
potential corresponding to the band offset $V_0$ at the A/B
heterointerface ($z=a$). The form of kinetic energy operator [the
first term in Eq.~(\ref{Hamiltonian})] corresponds to above BCs
(\ref{BCs})~\cite{BenDaniel1966}. Consider the eigenstates of
electron in QW without additional heterobarrier ($V_0=0$) and with
equal masses in layers A and B, $m_A=m_B$, as a basis set,
$\psi^0_n(z)e^{i(k_xx+k_yy)}$, where
$\psi^0_n(z)=(2/L)^{1/2}\sin(\pi nz/L)$. The corresponding
eigenenergies are well-known:
$E^0_n(k)=E_0n^2+\frac{\hbar^2k^2}{2m_A}$. Since we use the basis
set being the eigenstates of the Hamiltonian
$H_0=-\frac{\hbar^2\nabla^2}{2m_A}+U(z)$, then it is convenient to
extract the remaining part of the operator, $\Delta H=H-H_0$.

The translational invariance of the Hamiltonian in the QW plane is
conserved in the case of SQW as well. This means, that the
corresponding momentum operators can be replaced by its eigenvalues,
$p_x^2+p_y^2\rightarrow\hbar^2k^2$. The variables are not separated
due to the $z$-dependence of the effective mass. Thus, the spectral
problem is one-dimensional, but has a parametric dependence on $k$.
The additional term in the Hamiltonian is given by
\begin{eqnarray}
\label{DeltaH} \nonumber \Delta
H&=&V(z)-\frac{\hbar^2}{2}\frac{\partial m^{-1}}{\partial
z}\frac{\partial}{\partial
z}\\
&-&\frac{\hbar^2}{2}\left[\frac{1}{m_A}-\frac{1}{m(z)}\right]\left(k^2-\frac{\partial^2}{\partial
z^2}\right).
\end{eqnarray}

Now the exact step-like dependences of $V(z)$ and $m^{-1}(z)$ (see
Fig.~\ref{fig01}a,b) can be taken into account. The matrix elements
of the Hamiltonian $\langle n'|H|n\rangle\equiv H_{n'n}(k)$ can be
found analitically, $H_{n'n}(k)=E^0_n(k)\delta_{n'n}+(\Delta
H)_{n'n}(k)$~\cite{Note1}. The numerical diagonalization of the
Hamiltonian matrix is possible in the case of its finite size. The
truncation procedure is necessary: the choice of maximal $n$ is
determined by required accuracy of calculation. In the numerical
calculations, the value $n_{max}=500$ is used, which ensures perfect
accuracy and agreement with the result obtained by means of the
BC-based approach. The numerical diagonalization allows to obtain
the subband spectrum $E_i(k)$ and the corresponding expansion
coefficients $C^i_n$ relative to the above-mentioned basis set.
Thus, the corresponding transverse wave function has the form
$\psi_i(z;k)=\sum_nC^i_n(k)\psi^0_n(z)$.

\begin{figure}
\includegraphics{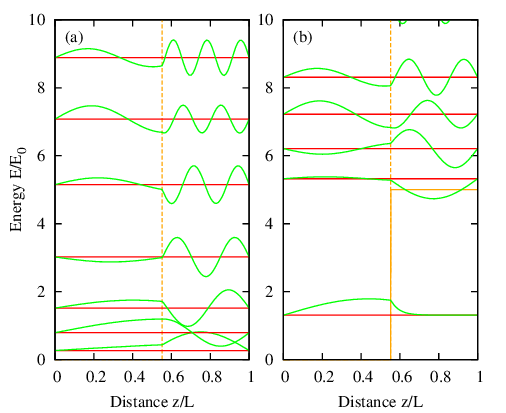}
\caption{\label{fig02} Subband energies $E_i(0)$ (solid red lines)
and corresponding wave functions (solid green line) for electron in
SQW. The heterointerface position $a/L=1-1/\sqrt{5}\approx 0.553$ is
marked by vertical dashed line, $m_B/m_A=15$. (a) No band offset,
$V_0=0$; (b) The potential $V(z)$ is indicated by solid orange line,
$V_0=5E_0$.}
\end{figure}

The results of the numerical diagonalization of the Hamiltonian
$H=H_0+\Delta H$ (at $k=0$) is presented in Fig.~\ref{fig02}. The
corresponding wave functions are shown as well. In the case $V_0=0$
(Fig.~\ref{fig02}a) one can see, that the wave functions are
predominantly localized in the region with a large effective mass.
The nodes of the wave function are shifted to the region B. In the
case $V_0\neq 0$ (Fig.~\ref{fig02}b) the same applies to the states
above the barrier [$E_n(0)>V_0$], whereas the ground state
[$E_1(0)<V_0$] is predominantly localized in region A. Thus, the
spatial region of the carrier localization in a heterostructure is
determined not only by the barrier height (band offset or
``potential'' energy), but by the total energy, which includes the
``kinetic'' part, parameterized by an inverse effective mass. In
addition, there is a possibility of localization of two electronic
states in different spatial regions within one structure: the ground
state is predominantly localized in region A, and the excited state
is in region B (see Fig.~\ref{fig02}b). This could be useful for
applications such as the design of quantum cascade lasers.

\begin{figure}
\includegraphics{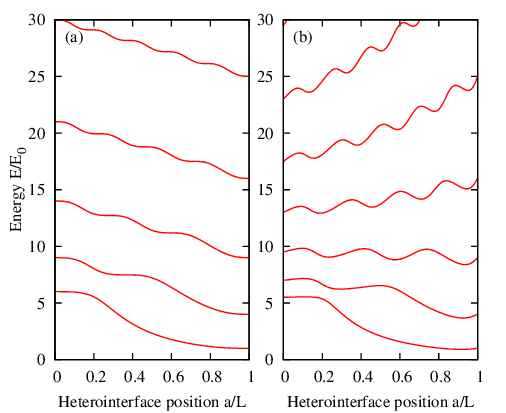}
\caption{\label{fig03} Dependences of the subband energies $E_i(0)$
on the heterointerface position $a$ ($0<a<L$). (a) Equal effective
masses, $m_A=m_B$. (b) Non-equal effective masses, $m_B/m_A=2$. The
band offset $V_0$ is the same in both cases, $V_0=5E_0$.}
\end{figure}

Let us consider the dependence of subband energies on the
hetereinteface position $z=a$ as additional confirmation of the
effect of the total energy on the quantum confinement. The
dependences of the energies $E_i(0)$ on $a$ are shown in
Fig.~\ref{fig03}. The cases $m_A=m_B$ and $m_A<m_B$ are reproduced
in Fig.~\ref{fig03}a and Fig.~\ref{fig03}b, respectively. The
barrier height $V_0$ is finite in both cases. In the first case, the
dependences $E_i(a)$ are monotonic. Similar behavior takes place at
$m_A>m_B$. Increasing the barrier width (decreasing $a$) obviously
leads to an upward shift of the levels. However, in the case
$m_A<m_B$, the $E_i(a)$ dependences can be non-monotonic, which
demonstrates that sometimes the ``kinetic'' contribution outweighs
the ``potential'' one.

In Fig.~\ref{fig03} $a=L$ corresponds to the absence of an
additional barrier, whereas at $a=0$ the barrier fills the entire QW
region, i.e., there is no barrier, but all levels are shifted upward
by $V_0$. In both cases, the spectral problem has an analytical
solution. The subband energies for $a=L$ and $a=0$ are given by
\begin{subequations}
\label{E_0}
\begin{align}
E_n=E_0n^2,\\
E_n=V_0+\frac{m_A}{m_B}E_0n^2,
\end{align}
\end{subequations}
respectively.

One can see, that, for fixed ratio $m_B/m_A>1$ and barrier height
$V_0$, for subbands with $n$ exceeding
$n_{th}=\sqrt{\frac{V_0}{E_0(1-m_A/m_B)}}$, the $n$-th subband at
$a=L$ has a higher energy than at $a=0$ (see Fig.~\ref{fig03}b).
Thus, the presence of regions of variation of $a$, where
$dE_i/da>0$, is inevitable when $m_B>m_A$. This behavior is typical
for levels lying above $V_0$, or lower energies in structures with
thin barriers [$(L-a)\ll L$].

Thus, the present calculation indicates a new possibility to control
the region of charge carrier localization in semiconductor
heterostructures. This mechanism can find application in
heterostructures with a large number of heterointerfaces, such as
quantum cascade lasers. The following advantages of used matrix
approach should be noted: i) it allows to take into account any
finite number of heteroboundaries (in a similar situation the number
of equations in the system for determining the spectrum, increases
significantly, whereas when using the matrix approach, only the
matrix elements change); ii) the approach can be generalized to
determine the hole spectrum and any many-band consideration (cf.
Refs.\cite{Kokurin2020} and \cite{Kokurin2023}); iii) the smooth
variation of the band offset $V(z)$ and effective mass $m(z)$, as in
the graded-gap semiconductor, can be taken into account, while the
BC-based approach is applicable only to sharp heterojunctions.

In conclusion, using a simple SQW model, a new possibility for
controlling quantum confinement in semiconductor heterostructures is
demonstrated. The importance of the total energy in heterolayers is
shown, i.e., the quantum confinement is determined not only by the
barrier height (the band offset at the heterointerfaces), but also
by effective mass or other band parameters. The proposed technique
can serve as an additional tool for adjusting energy levels
(subbands) in semiconductor structures.

We carried out calculations without reference to specific materials
and their band structure parameters. The presented considerations
seem somewhat overdone, especially if we turn to conventional
heterostructures such as GaAs/AlGaAs (where the effective mass ratio
is of the order of 1.3). However, even if suitable heteropairs with
a large effective mass ratio do not yet exist, in today's reality,
when new materials are discovered and synthesized daily, it seems
possible to create them purposefully. The existence of materials
with so-called flat bands (the effective mass is very large) seems
useful. The above consideration can also be generalized to the case
of non-planar heterointerfaces, e.g., in spherical core-shell
nanocrystals.

This research was supported by the Russian Science Foundation (Grant
No. 21-72-30020-$\Pi$, https://rscf.ru/project/21-72-30020/).

\clearpage

\setcounter{page}{1} \setcounter{equation}{0} \setcounter{figure}{0}
\setcounter{table}{0} \setcounter{section}{0}

\renewcommand{\thepage}{S\arabic{page}}
\renewcommand{\theequation}{S\arabic{equation}}
\renewcommand{\thefigure}{S\arabic{figure}}
\renewcommand{\thetable}{S\arabic{table}}
\renewcommand{\thesection}{S\arabic{section}}

\title{Supplemental materials for ``On quantum confinement without barriers''}

\begin{center}
    \textbf{\large Supplemental materials for ``On quantum confinement without barriers''} \\
\end{center}
I.A. Kokurin\\
Ioffe Institute, 194021 St. Petersburg, Russia

\section{SQW Hamiltonian and its matrix elements}

The Hamiltonian for an electron with scalar $z$-dependent effective
mass $m(z)$ in structured quantum well (SQW) is given by
\begin{equation}
H=-\frac{\hbar^2}{2}\nabla\frac{1}{m(z)}\nabla+U(z)+V(z).
\end{equation}
Here $U(z)$ is the hard-wall potential corresponding to zero BCs at
$a=0,L$, $V(z)$ is the additional potential step $V_0$ at the A/B
heterointerface.

Let us use the eigenfunctions
\begin{equation}
\psi^0_{nk}({\bf r})=\sqrt{\frac{2}{L}}\sin\left(\frac{\pi
n}{L}z\right)e^{i(k_xx+k_yy)}
\end{equation}
of the Hamiltonian $H_0=-\frac{\hbar^2\nabla^2}{2m_A}+U(z)$ as the
basis set. Corresponding eigenenergies have the form
\begin{equation}
E^0_n(k)=E_0n^2+\frac{\hbar^2k^2}{2m_A},
\end{equation}
where $E_0=\frac{\pi^2\hbar^2}{2m_AL^2}$.

It is convenient to extract the remaining part of the total
Hamiltonian, $\Delta H=H-H_0$
\begin{eqnarray}
\nonumber \Delta H&=&V(z)-\frac{\hbar^2}{2}\frac{\partial
m^{-1}}{\partial z}\frac{\partial}{\partial
z}\\
&-&\frac{\hbar^2}{2}\left[\frac{1}{m_A}-\frac{1}{m(z)}\right]\left(k^2-\frac{\partial^2}{\partial
z^2}\right).
\end{eqnarray}
It should be noted, that this term does not violate translation
invariance in QW plane. At the same time, the variables in
Schr\"{o}dinger equation are not separated due to the spatial
dependence of effective mass. However, the spectral problem is
reduced to 1-dimensional one with parametric dependence on in-plane
momentum $\hbar k$.

If we use the following step-like dependences of effective mass
\begin{equation}
\frac{1}{m(z)}=\left\{
\begin{array}{l}
m^{-1}_A\;\; \text{if}\;\; 0<z\leq a\\
m^{-1}_B\;\; \text{if}\;\; a<z\leq L
\end{array} \right.,
\end{equation}
and heteropotential
\begin{equation}
V(z)=\left\{
\begin{array}{l}
0 \;\;\;\; \text{if}\;\; 0<z\leq a\\
V_0\;\;\text{if}\;\; a<z\leq L
\end{array} \right.,
\end{equation}
then the operator $\Delta H$ is transformed into the form
\begin{eqnarray}
\nonumber \Delta
H&=&\frac{\hbar^2}{2m_A}\left(1-\frac{m_A}{m_B}\right)\delta(z-a)\frac{\partial}{\partial
z}\\
&+&\left\{
\begin{array}{l}
0,\qquad\qquad\qquad\qquad\;\;\;\;\; \text{if} \;\; 0<z\leq a\\
V_0-E^0_n(k)\left(1-\frac{m_A}{m_B}\right),\; \text{if} \;\; a<z\leq
L
\end{array}\right.
\end{eqnarray}

The step-like dependences $m^{-1}(z)$ and $V(z)$ leads to the
necessity to find the overlap integrals of basis functions
$I_{n'n}=\int_a^Ldz\psi^0_{n'}(z)\psi^0_n(z)$ in the barrier region.
The simple calculations lead to the following expressions
\[
I_{n'n}=\frac{\sin[\pi(n'+n)a/L]}{\pi(n'+n)}-\frac{\sin[\pi(n'-n)a/L]}{\pi(n'-n)},
\]
\[
I_{nn}=\left(1-\frac{a}{L}\right)+\frac{\sin[2\pi na/L]}{2\pi n}.
\]

The operator $\Delta H$ can be divided into three parts, $\Delta
H=\Delta H_1+\Delta H_2+\Delta H_3$. The corresponding matrix
elements are given by
\begin{equation}
(\Delta
H_1)_{n'n}=E_0\frac{2n}{\pi}\left(1-\frac{m_A}{m_B}\right)\sin\left(\frac{\pi
n'a}{L}\right)\cos\left(\frac{\pi na}{L}\right),
\end{equation}

\begin{equation}
(\Delta H_2)_{n'n}=V_0I_{n'n},
\end{equation}
and
\begin{equation}
(\Delta H_3)_{n'n}=-E^0_n(k)\left(1-\frac{m_A}{m_B}\right)I_{n'n}.
\end{equation}

Thus, the spectral problem comes down to the numerical
diagonalization of truncated matrix ($n_{max}=500$) with elements
$H_{n'n}(k)=E^0_n(k)\delta_{n'n}+(\Delta H)_{n'n}(k)$. The result is
the subband energy spectrum $E_i(k)$ ($i=1,2,3,...$) and the
expansion coefficients $C^i_n(k)$. The latter allow to write down
the corresponding transverse wave functions:
\begin{equation}
\psi_i(z;k)=\sum_nC^i_n(k)\psi^0_n(z),
\end{equation}
where $\psi^0_n(z)=(2/L)^{1/2}\sin(\pi nz/L)$.


\begin{thebibliography}{8}

\bibitem{Bastard1988} G. Bastard, {\it Wave Mechanics Applied to Semiconductor
Heterostructures} (Les Editions des Physique, Les Ulis, 1988).

\bibitem{Vurgaftman2001} I. Vurgaftman, J. R. Meyer, and L. R. Ram-Mohan, J. Appl. Phys.
{\bf 89}, 5815 (2001).

\bibitem{BenDaniel1966} D. J. BenDaniel and C. B. Duke, Space-Charge Effects on Electron
Tunneling, Phys. Rev. {\bf 152}, 683 (1966).

\bibitem{Rudakov2019} A. O. Rudakov, I. A. Kokurin, Electronic States in Cylindrical Core-Multi-Shell
Nanowire. Semiconductors {\bf 53}, 2137 (2019).

\bibitem{Note1} See Supplemental Materials at ???? for the detailed form of
  the matrix Hamiltonian.

\bibitem{Kokurin2020} I. A. Kokurin, Electronic States in Nanowires
with Hexagonal Cross-Section. Semiconductors {\bf 54}, 1897 (2020).

\bibitem{Kokurin2023} I. A. Kokurin, Dimensional quantization and zero-field spin
splitting of holes in GaAs nanowires. Phys. Rev. B {\bf 108}, 165301
(2023).

\end{thebibliography}
\end{document}